\documentclass[aps,prb,reprint,showkeys,longbibliography]{revtex4-1}
\usepackage{amsmath}
\usepackage{amssymb}
\usepackage{amsfonts}
\usepackage{graphicx}
\usepackage{bm}
\usepackage{color}

\begin{document}

\title{Experimental observation of third-order effect in magnetic small-angle neutron scattering}

\author{Konstantin L. Metlov}\email[Electronic address: ]{metlov@fti.dn.ua}
\affiliation{Donetsk Institute for Physics and Technology, R.\ Luxembourg str.\ 72, 83114~Donetsk, Ukraine}
\affiliation{Institute for Numerical Mathematics RAS, 8~Gubkina str., 119991~Moscow GSP-1, Russia}
\author{Kiyonori~Suzuki}
\affiliation{Department of Materials Science and Engineering, Monash University, Clayton, Victoria~3800, Australia}
\author{Dirk Honecker}
\affiliation{Department of Physics and Materials Science, University of Luxembourg, 162A~Avenue de la Fa\"iencerie, L-1511~Luxembourg, Grand Duchy of Luxembourg}
\author{Andreas Michels}\email[Electronic address: ]{andreas.michels@uni.lu}
\affiliation{Department of Physics and Materials Science, University of Luxembourg, 162A~Avenue de la Fa\"iencerie, L-1511~Luxembourg, Grand Duchy of Luxembourg}

\date{\today}

\begin{abstract}
A recent theory [Metlov and Michels, Phys.\ Rev.\ B~$\mathbf{91}$, 054404 (2015)] predicts a qualitatively new effect in the magnetic small-angle neutron scattering (SANS) cross section of statistically-isotropic disordered ferromagnetic media. The effect is due to the third-order terms in the amplitude of the inhomogeneities. Here, its existence is demonstrated both numerically via large-scale micromagnetic simulations and analyzed experimentally in a two-phase iron-based nanocomposite. The previous model is extended to an arbitrary spatial defect profile, which allows us to describe the experimental field dependence of the third-order SANS effect quantitatively.
\end{abstract}

\keywords{neutron scattering; small-angle neutron scattering; magnetic materials; micromagnetism; noncollinear magnetic structures}

\maketitle\

\section{Introduction}

Magnetic small-angle neutron scattering (SANS) is a powerful tool for investigating nonuniform magnetization structures on a mesoscopic length scale ($\sim 1-300 \, \mathrm{nm}$) inside magnetic materials (see Ref.~\cite{rmp2019} for a recent review). An advantage of the SANS technique, compared e.g.\ to electron-microscopy-based methods, is that it provides statistically-averaged information about a large number of scattering objects. When conventional SANS is supplemented by ultra or very small-angle neutron scattering the spatial resolution can be extended up to the micrometer range~\cite{jericha2012,jericha2013}. This is an important size regime in which many macroscopic material properties are realized. Magnetic SANS has previously been applied to study the spin structures of a wide range of materials such as magnetic nanoparticles~\cite{disch2012,guenther2014,maurer2014,bender2015,grutter2017,bender2018jpcc,bender2018prb,oberdick2018,krycka2019,benderapl2019,bersweiler2019,zakutna2020}, hard and soft magnetic nanocomposites~\cite{suzuki2007,herr08pss,lister2010,ono2016}, proton domains~\cite{michels06a,aswal08nim,noda2016}, magnetic steels~\cite{bischof07,bergner2013,Pareja:15,shu2018}, reentrant spin glasses~\cite{mirebeau2018}, or Heusler-type alloys~\cite{bhatti2012,runov2006,michelsheusler2019,leighton2019}.

In Ref.~\onlinecite{metmi2015}, based on the analytical solution of the corresponding micromagnetic problem, we have derived expressions for the SANS cross section of a ferromagnetic medium with a weakly inhomogeneous uniaxial magnetic anisotropy, saturation magnetization, and exchange stiffness, which is valid up to the third order in the (small) amplitudes of the inhomogeneities. It follows e.g.\ that the second-order SANS cross section at sufficiently small values of the applied magnetic field inevitably displays a prominent UFO-like shape~\cite{perigo2014}. For periodic systems of defects, this theory also reproduces magnetic configurations (see Fig.~5 in Ref.~\onlinecite{michelsdmi2019}), in which the positions of vortices and saddles satisfy certain constraints~\cite{BM18,BM17}, whose topological origins can be traced back to Abel's theorem. But the central result of Ref.~\onlinecite{metmi2015} is that under very general assumptions regarding the type, distribution, and magnitude of random inhomogeneities of material parameters in a magnet, a specific combination of SANS cross-section values as a function of the scattering vector length is exactly zero in the second order, whereas the third-order contribution to this combination is nonzero and has a nontrivial functional dependence on the scattering vector, magnetic field, and the average exchange length.

Normally, the higher-order contributions are insignificant as compared to the larger, lower-order ones. But the cancellation of the second-order terms allows one to unmask the third-order effect and opens it for direct experimental observation and analysis, which is the main purpose of the present work.

In Sec.~\ref{msanscrosssection} we provide the basic equations for the magnetic SANS cross section in the perpendicular scattering geometry. Sections~\ref{msanssecondorder} and \ref{msansthirdorder} then briefly summarize the expressions for the second and higher-order terms in the SANS cross sections. Our numerical and experimental results are presented and discussed in Sec.~\ref{results}. This section also includes the expression for the third-order effect function for arbitrary spatial defect profile and our analysis of the experimental data.

\section{Summary of previous results}

\subsection{Magnetic SANS Cross Section}
\label{msanscrosssection}

\begin{figure}[tb!]
\centering
\resizebox{1.0\columnwidth}{!}{\includegraphics{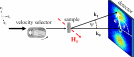}}
\caption{\label{fig:setup}Schematic drawing of the SANS setup (see main text for explanations).}
\end{figure}

Magnetic SANS experiments are commonly performed in a setup schematically shown in Fig.~\ref{fig:setup}.
The experiment measures the scattering cross section as a function of the scattering vector $\mathbf{q} = \mathbf{k}_1 - \mathbf{k}_0$, being the difference between the wave vectors of the scattered ($\mathbf{k}_1$) and incident ($\mathbf{k}_0$) neutrons; its magnitude $q = |\mathbf{q}| = (4\pi/\lambda) \sin(\psi/2)$ depends on the mean wavelength $\lambda$ of the neutrons (selected by the velocity selector) and on the scattering angle $\psi$. The applied-field direction $\mathbf{H}_0$ is parallel to the $\mathbf{e}_z$-direction of a Cartesian laboratory coordinate system and perpendicular to the incident neutron beam ($\mathbf{k}_0 \parallel \mathbf{e}_x \perp \mathbf{H}_0$). In the small-angle approximation, the component of $\mathbf{q}$ along $\mathbf{k}_0$ is neglected, i.e., $\mathbf{q} \cong \{ 0, q_y, q_z \} = q \{ 0, \sin\theta, \cos\theta \}$, where the angle $\theta$ specifies the orientation of $\mathbf{q}$ on the two-dimensional detector.

It is well known that the discrete atomic structure of matter is generally of no relevance for small-angle scattering. The cross sections are therefore expressed in terms of suitably coarse-grained continuum variables, represented by their Fourier transforms. The latter are denoted here by a tilde over the symbol, so that for the nuclear scattering-length density we have $\widetilde{N}(\mathbf{q})=\iiint N(\mathbf{r})e^{-\imath \mathbf{q}\mathbf{r}} d V$. Similarly, the Fourier image of the coordinate-dependent saturation magnetization of the material $M_s(\mathbf{r})$ is denoted as $\widetilde{M}_s(\mathbf{q})$, and $\widetilde{\mathbf{M}}(\mathbf{q}) = \{ \widetilde{M}_x(\mathbf{q}), \widetilde{M}_y(\mathbf{q}), \widetilde{M}_z(\mathbf{q}) \}$ is the Fourier transform of the magnetization vector field $\mathbf{M}(\mathbf{r}) = \{ M_x(\mathbf{r}), M_y(\mathbf{r}), M_z(\mathbf{r}) \}$. Then, the total unpolarized elastic SANS cross section $d \Sigma / d \Omega$ can be written as~\cite{rmp2019}:
\begin{equation}
\label{sigmasansperp2d}
\frac{d \Sigma}{d \Omega}(\mathbf{q}) = \frac{d \Sigma_{\mathrm{res}}}{d \Omega}(\mathbf{q}) + \frac{d \Sigma_{\mathrm{SM}}}{d \Omega}(\mathbf{q}) ,
\end{equation}
where
\begin{equation}
\label{sigmaresperp0}
\frac{d \Sigma_{\mathrm{res}}}{d \Omega}(\mathbf{q}) = \frac{8 \pi^3}{V} \left( |\widetilde{N}|^2 + b_H^2 |\widetilde{M}_s|^2 \sin^2\theta \right)
\end{equation}
represents the nuclear and magnetic residual SANS cross section, which is measured at complete magnetic saturation (large applied field), and the remaining term
\begin{eqnarray}
\label{sigmasmperp0}
\frac{d \Sigma_{\mathrm{SM}}}{d \Omega}(\mathbf{q}) = \frac{8 \pi^3}{V} b_H^2 \left( |\widetilde{M}_x|^2 + |\widetilde{M}_y|^2 \cos^2\theta \right. \nonumber \\ \left. + \left[ |\widetilde{M}_z|^2 - |\widetilde{M}_s|^2 \right] \sin^2\theta \right. \nonumber \\ \left. - (\widetilde{M}_y \widetilde{M}_z^{\ast} + \widetilde{M}_y^{\ast} \widetilde{M}_z) \sin\theta \cos\theta \right) ,
\end{eqnarray}
denotes the spin-misalignment SANS cross section, which vanishes at saturation when the real-space magnetization is given by $\mathbf{M} = \{0, 0, M_z = M_s(\mathbf{r}) \}$. In the preceding expressions $V$ is the scattering volume and $b_H = 2.91 \times 10^8 \, \mathrm{A}^{-1}\mathrm{m}^{-1}$ is the magnetic scattering length in the small-angle regime (the atomic magnetic form factor is approximated by $1$, since we are dealing with forward scattering). Note that Ref.~\onlinecite{metmi2015} uses a different definition of the Fourier transform, so that there appears an insubstantial difference in the prefactors.

\subsection{Second-order magnetic SANS}
\label{msanssecondorder}

The Fourier images of the magnetization components $\widetilde{\mathbf{M}}(\mathbf{q})$ for a particular material (class of materials) at different values of the applied field can be found by solving the corresponding micromagnetic problem. When the material is uniform and boundless, its magnetization will also be uniform under any nonzero applied field. For small deviations from uniformity, it is possible to build the analytical solution of the micromagnetic problem for $\widetilde{\mathbf{M}}(\mathbf{q})$ on top of the well-known theory of the approach-to-magnetic saturation~\cite{schloemann67,schloemann71}. The chief difference is that the latter is concerned with the value of the average magnetization $\widetilde{\mathbf{M}}(0)$, whereas the magnetic SANS cross section (\ref{sigmasmperp0}) depends on all its Fourier harmonics. Yet, the setting of the micromagnetic problem is very similar. It is assumed that the local saturation magnetization is a function of the position $\mathbf{r} = \left\{ x, y, z \right\}$ inside the material:
\begin{equation}
\label{msatdef}
M_s(\mathbf{r}) = M_s [1 + I_m(\mathbf{r})] ,
\end{equation}
where $I_m$ is an inhomogeneity function, small in magnitude, which describes the local variation of $M_s(\mathbf{r})$. Similar spatial inhomogeneities can be present in the magnetostatic exchange length $l_M^2(\mathbf{r}) = 2 A(\mathbf{r})/[\mu_0 M_s^2(\mathbf{r})] = l_0^2 [1 + I_e(\mathbf{r})]$ and in the dimensionless quality factor $Q(\mathbf{r}) = 2K(\mathbf{r})/[\mu_0 M_s^2(\mathbf{r})] = Q_0 I_k(\mathbf{r})$, where $A$ and $K$ are the spatially-dependent exchange stiffness and the uniaxial anisotropy constant, respectively. The spatial averages of the inhomogeneity functions are assumed to be zero:~$\langle I_{m,e,k}(\mathbf{r}) \rangle = 0$. Consequently, $\langle M_s(\mathbf{r}) \rangle = M_s$ is the average saturation magnetization of the sample, which can be measured with a magnetometer.

Assuming that the inhomogeneity functions are small quantities $I_{m,e,k} \ll 1$ of the same order, the solution of the micromagnetic problem can be expressed as a Taylor series:
\begin{equation}
\widetilde{\mathbf{M}} = \{ 0, 0, M_s \} \delta(\mathbf{q}) + \widetilde{\mathbf{M}}^{(1)} + \widetilde{\mathbf{M}}^{(2)} + \ldots ,
\label{taylorexpansionsat}
\end{equation}
where $\delta(\mathbf{q})$ is the Dirac's delta function, and $\widetilde{\mathbf{M}}^{(i)}$ contains the terms of the order $i$ in $I_{m,e,k}(\mathbf{r})$. The first term in Eq.~(\ref{taylorexpansionsat}) corresponds to the saturated state. Solving the micromagnetic problem up to the first order and obtaining expressions for $\widetilde{\mathbf{M}}^{(1)}$ allows one to express the magnetic SANS cross section up to the second order because Eq.~(\ref{sigmasmperp0}) is quadratic in $\widetilde{\mathbf{M}}$. The first set of such expressions was obtained in~\cite{michels2013} for the case of inhomogeneous saturation magnetization and anisotropy. They were verified and extended in many follow-up works (including Ref.~\onlinecite{metmi2015}) and admit a very convenient representation:
\begin{eqnarray}
\label{sigmasmperp}
\frac{d \Sigma_{\mathrm{SM}}}{d \Omega}(\mathbf{q}) = S_H(\mathbf{q}) R_H(\mathbf{q}, H_i) + S_M(\mathbf{q}) R_M(\mathbf{q}, H_i) ,
\end{eqnarray}
where the contribution $S_H R_H$ is due to perturbing magnetic anisotropy fields and the part $S_M R_M$ is related to magnetostatic fields; $H_i$ is the internal magnetic field, consisting of $H_0$ and of the average demagnetizing field due to the shape of the sample. The anisotropy-field scattering function
\begin{equation}
\label{shdef}
S_H(\mathbf{q}) = \frac{8 \pi^3}{V} b_H^2 |\mathbf{\widetilde{H}}_p|^2
\end{equation}
depends on the Fourier transform $\mathbf{\widetilde{H}}_p(\mathbf{q})$ of the local magnetic anisotropy field, whereas the scattering function of the longitudinal magnetization
\begin{equation}
\label{smdef}
S_M(\mathbf{q}) = \frac{8 \pi^3}{V} b_H^2 |\widetilde{M}_z|^2
\end{equation}
characterizes the spatial variations of the saturation magnetization. The latter can be related to the mean magnitude $\Delta M \propto \widetilde{M}_z$ of the magnetization jump at internal (e.g., particle-matrix) interfaces. The dimensionless micromagnetic response functions are given by:
\begin{equation}
\label{rhdefperp}
R_H(q, \theta, H_i) = \frac{p^2}{2} \left( 1 + \frac{\cos^2\theta}{\left( 1 + p \sin^2\theta \right)^2} \right)
\end{equation}
and
\begin{equation}
\label{rmdefperp}
R_M(q, \theta, H_i) = \frac{p^2 \sin^2\theta \cos^4\theta}{\left( 1 + p \sin^2\theta \right)^2} + \frac{2 p \sin^2\theta \cos^2\theta}{1 + p \sin^2\theta} ,
\end{equation}
where $p = 1/(h + l_M^2 q^2)$, $h = H_i/M_s$, and the magnetostatic exchange length equals $l_M \sim 3-10 \, \mathrm{nm}$ (Ref.~\onlinecite{kronfahn03}). Alternatively, the function $p = M_s/[H_i ( 1 + l_H^2 q^2 )]$ can be expressed via the micromagnetic exchange length $l_H(H_i) = \sqrt{2 A/(\mu_0 M_s H_i)}$, which characterizes the range over which perturbations in the spin structure decay~\cite{michels03prl,michelsprb2010,michels2013}

We emphasize that it is $d \Sigma_{\mathrm{SM}} / d \Omega$ which depends on the magnetic interactions (exchange, anisotropy, magnetostatics, etc.), while $d \Sigma_{\mathrm{res}} / d \Omega$ is determined by the geometry of the underlying grain microstructure (e.g., the particle shape or the particle-size distribution). One way to access the magnetic interactions is to subtract the residual scattering cross section measured at a large saturating field from the total $d \Sigma / d \Omega$ at a lower field. This is not always possible in experimental situations because of the difficulty to achieve complete magnetic saturation of the sample. The other approach~\cite{michels2013} is to use the bilinearity of Eq.~(\ref{sigmasmperp}) in $R_H$ and $R_M$, which are simple functions of $\mathbf{q}$, $H_0$, and $l_M$ only. Linear regression allows then to compute $S_M$, $S_H$, and by extrapolation $d \Sigma_{\mathrm{res}} / d \Omega$ at each $q$ without the necessity to magnetically saturate the sample. Analyzing in this way azimuthally-averaged SANS cross sections at different fields as functions of the magnitude of the scattering vector is a reliable and very precise method~\cite{michels2013} for obtaining the value of the exchange-stiffness constant $A$.

\subsection{Third-order effect in magnetic SANS}
\label{msansthirdorder}

Normally, the higher-order effects are masked by the lower-order ones. But in magnetic SANS the third-order contribution can be unmasked by considering the following combination of the perpendicular unpolarized cross-section values~\cite{metmi2015}:
\begin{eqnarray}
\Delta\Sigma_{\mathrm{SM}} = \left. \frac{d \Sigma_{\mathrm{SM}}}{d \Omega} \right|_{\theta = 0} - 2 \left. \frac{d \Sigma_{\mathrm{SM}}}{d \Omega}\right|_{\theta = \pi/2}.
\label{eq:deltasigma} 
\end{eqnarray}
The second-order contribution in $\Delta\Sigma_{\mathrm{SM}}$ is exactly zero, which can also be seen from Eqs.~(\ref{sigmasmperp})$-$(\ref{rmdefperp}). This cancellation is a universal property of the SANS cross section from disordered ferromagnets, independent of the specific spatial profile of the defects. Assuming for simplicity that the inhomogeneity functions are related via $I_m = I$ and $I_k = \kappa I$ with $\kappa \lesssim 1$, the remaining third-order contribution is nonzero and takes on an especially simple form~\cite{metmi2015}:
\begin{eqnarray}
\frac{\Delta\Sigma_\mathrm{SM}V}{32 \pi^3 b_H^2} =
\left. \left\langle \frac{\widetilde{M}_x^{(1)}\otimes \widetilde{M}_x^{(1)}+\widetilde{M}_y^{(1)}\otimes \widetilde{M}_y^{(1)}}{2} \widetilde{I} \right\rangle \right|_{\genfrac{}{}{0pt}{}{q_z=0}{q_x = 0}} ,
\label{thirdorderintermediate}
\end{eqnarray}
where $\otimes$ denotes the discrete convolution in $\mathbf{q}$-space $q = q_y$, and the angular brackets denote a triple (configurational, directional, and anisotropy direction) average. Each of the $\widetilde{M}_x^{(1)}$ and $\widetilde{M}_y^{(1)}$ is proportional to $\widetilde{I}$ in the first order, so that their product multiplied by $\widetilde{I}$ is of the third order in $\widetilde{I}$. Note that in Ref.~\cite{metmi2015} discrete Fourier transforms were used with the dimensions of $\widetilde{M}_i$ and $M_i$ being the same. For spherical Gaussian defects with a spatial defect profile $\propto e^{-r^2/s^2}$, where $s$ is the defect size, it is possible to split the $\kappa$-dependent terms and obtain the following expression for $\Delta\Sigma_{\mathrm{SM}}$~\cite{metmi2015}:
\begin{eqnarray}
\Delta\Sigma_{\mathrm{SM}} \propto \kappa^2 g_{\mathrm{A}}(\mu,h,\lambda) + g_{\mathrm{MS}}(\mu, h,\lambda),
\label{eq:sigmaperpthree}
\end{eqnarray}
where the dimensionless functions $g_{\mathrm{A}}$ and $g_{\mathrm{MS}}$ depend on the reduced scattering vector $\mu = q s$, the reduced magnetic field $h$, and on the dimensionless parameter $\lambda = l_M / s$. These functions are plotted in Figs.~4 and 5 of Ref.~\onlinecite{metmi2015}. 

Now we have all the tools at hand to address the main questions of this paper: 1)~whether the third-order magnetic SANS effect can be detected in experimental data and numerical micromagnetic simulations, so that the difference $\Delta\Sigma_{\mathrm{SM}}$ is nonzero; and 2)~whether the measured $\Delta\Sigma_{\mathrm{SM}}$ can indeed be described by Eqs.~(\ref{thirdorderintermediate}) and (\ref{eq:sigmaperpthree}).

\section{Results and discussion}
\label{results}

First, to confirm the existence of the third-order effect, we have performed
micromagnetic simulations of the magnetic SANS cross section in an artificial system of magnetic holes. These numerical computations (see Ref.~\onlinecite{erokhin2015} for details) were adapted to the microstructure of porous iron with a volume fraction of $32 \, \%$ and with randomly placed pore centers. The simulation code takes into account the four standard contributions to the total magnetic energy, i.e., energy in the external magnetic field, magnetic anisotropy, exchange and dipolar interaction energies. The sample volume $V = 0.2 \times 0.75 \times 0.75 \, \mathrm{\mu m}^3$ was divided into $N \sim 5 \times 10^5$ mesh elements, comprising both pores and nanocrystallites. Due to the flexibility of the mesh-generation algorithm, the shape of the pores can be controlled and was taken to be polyhedron-like. The pore-size distribution was assumed to be lognormal~\cite{krill98} with a median of $15 \, \mathrm{nm}$ and a variance of $1.16$, which yields a maximum of the distribution at $12 \, \mathrm{nm}$. The \textit{local} saturation magnetization of each Fe nanocrystallite was taken as $\mu_0 M_s = 2.2 \, \mathrm{T}$, which in conjunction with the above mentioned porosity value yields $\mu_0 \overline{M_s} \cong 1.5 \, \mathrm{T}$ for the entire sample. For the exchange-stiffness constant and the first cubic anisotropy constant of Fe, we have, respectively, assumed values of $A = 25 \, \mathrm{pJ/m}$ and $K_1 = 47 \, \mathrm{kJ/m^3}$ (Ref.~\onlinecite{cullitygraham05}). The direction of anisotropy axes varies randomly from crystallite to crystallite. The energy-minimization procedure provides (at some particular value of the applied magnetic field) the magnetization vector field $\mathbf{M}(\mathbf{r}) = \{ M_x(\mathbf{r}), M_y(\mathbf{r}), M_z(\mathbf{r}) \}$ of the sample on an \textit{irregular} lattice. This distribution is then mapped onto a \textit{regular} lattice, which permits us to calculate the magnetization Fourier coefficients and the ensuing neutron scattering cross section using Fast Fourier transformation. Further details can be found in Refs.~\onlinecite{erokhin2012prb,michels2012prb1,michels2014jmmm}.

Nuclear scattering was not considered and only the total magnetic SANS cross section [Eq.~(\ref{sigmasansperp2d})] without the nuclear term was computed. Numerical simulations are not limited by the series expansion and include the terms of all orders in the inhomogeneity amplitude.

\begin{figure}[tb!]
\centering{\includegraphics[width=1.0\columnwidth]{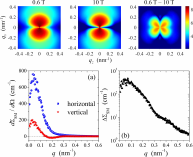}}
\caption{Results of micromagnetic simulations of nanoporous Fe for the third-order magnetic SANS effect~\cite{erokhin2015}. (upper row) Illustration of the subtraction procedure between the total $d \Sigma_{\mathrm{mag}} / d \Omega$ at $0.6 \, \mathrm{T}$ and at $10.0 \, \mathrm{T}$ (logarithmic color scale). (a)~Spin-misalignment SANS cross section along the horizontal and vertical directions (see inset). (b)~Resulting $\Delta\Sigma_{\mathrm{SM}}(q)$ computed according to Eq.~(\ref{eq:deltasigma}) (log-linear scale).}
\label{fig3}
\end{figure}

The spin-misalignment SANS cross section Eq.~(\ref{sigmasmperp0}) was obtained by taking the difference between the total magnetic $d \Sigma / d \Omega$ at $0.6 \, \mathrm{T}$ and at a larger magnetic field of $10 \, \mathrm{T}$, which approximates the magnetically saturated state. The resulting difference pattern exhibits the clover-leaf anisotropy with maxima roughly along the diagonals of the detector (upper row in Fig.~\ref{fig3}). This angular anisotropy is related to the dipolar fields which emerge from the jump of the magnetization magnitude at the pore-matrix interface~\cite{erokhin2015}. 

The quantity $\Delta\Sigma_{\mathrm{SM}}$ was computed by subtracting twice the spin-misalignment SANS cross section values along the vertical ($q_z = 0$) direction from its values along the horizontal ($q_y = 0$) direction. These curves (including the resulting $\Delta\Sigma_{\mathrm{SM}}$) are shown in Fig.~\ref{fig3}(a) and (b). The simulations yield nonzero values of $\Delta\Sigma_{\mathrm{SM}}$, which cannot be described by the second-order SANS theory. The plot of $\Delta\Sigma_{\mathrm{SM}}$ on a semi-logarithmic scale, shown in Fig.~\ref{fig3}(b), is mostly linear, which generally fits well with the prediction of Ref.~\onlinecite{metmi2015}.
\begin{figure*}[tb]
\centering{\includegraphics[width=1.0\textwidth]{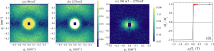}}
\caption{SANS results on NANOPERM [$(\mathrm{Fe}_{0.985}\mathrm{Co}_{0.015})_{89}\mathrm{Zr}_7\mathrm{B}_3$]. Total (nuclear and magnetic) unpolarized SANS cross section $d \Sigma / d \Omega$ in units of $100\, \mathrm{cm}^{-1} \mathrm{sr}^{-1}$ at (a)~$\mu_0 H_0 = 196 \, \mathrm{mT}$ and (b)~$\mu_0 H_0 = 1270 \, \mathrm{mT}$. (c)~Spin-misalignment SANS cross section $d \Sigma_{\mathrm{SM}} / d \Omega$ at $196 \, \mathrm{mT}$, i.e., $\frac{d \Sigma_{\mathrm{SM}}}{d \Omega}(196 \, \mathrm{mT}) = \frac{d \Sigma}{d \Omega}(196 \, \mathrm{mT}) - \frac{d \Sigma}{d \Omega}(1270 \, \mathrm{mT})$ (logarithmic color scale) ($\mathbf{k}_0 \perp \mathbf{H}_0$). The applied magnetic field $\mathbf{H}_0$ is horizontal in the plane. (d)~Solid line:~Normalized room-temperature magnetization curve. Data points ($1270 \, \mathrm{mT}$, $312 \, \mathrm{mT}$, $196 \, \mathrm{mT}$, $103 \, \mathrm{mT}$, $61 \, \mathrm{mT}$) specify the fields where the SANS measurements have been performed.}
\label{fig1}
\end{figure*}

To perform a quantitative analysis of the theoretical predictions for the third-order magnetic SANS effect we have used an existing experimental data set from the two-phase Fe-based alloy NANOPERM~\cite{suzuki06}. This material with a nominal composition of $(\mathrm{Fe}_{0.985}\mathrm{Co}_{0.015})_{89}\mathrm{Zr}_7\mathrm{B}_3$ consists of a dispersion of Fe nanoparticles, which are embedded in an amorphous magnetic matrix (particle size:~$15 \pm 2 \, \mathrm{nm}$; crystalline volume fraction:~$65 \, \%$; saturation magnetization:~$1.64 \, \mathrm{T}$). The raw SANS cross-section data for this material were already published in Ref.~\onlinecite{honecker2013}. The field-dependent azimuthally-averaged SANS cross sections can be excellently described by the second-order magnetic SANS theory (see Fig.~4(b) in \onlinecite{honecker2013}), yielding information on the magnetic interactions (exchange-stiffness constant, magnetostatic, and anisotropy fields).

Figure~\ref{fig1} shows the two-dimensional total unpolarized SANS cross section $d \Sigma / d \Omega$, computed from the raw data of Ref.~\onlinecite{honecker2013}, at an external field of $196 \, \mathrm{mT}$ [Fig.~\ref{fig1}(a)] and at a large value of the magnetic field of $1270 \, \mathrm{mT}$ [Fig.~\ref{fig1}(b)]. The experimental data set at $1270 \, \mathrm{mT}$ can be taken as an approximation to the residual SANS cross section $d \Sigma_{\mathrm{res}} / d \Omega$ [Eq.~(\ref{sigmaresperp0})], corresponding to the scattering signal in the completely saturated state [compare the hysteresis loop in Fig.~\ref{fig1}(d)]. It can be seen that the scattering at saturation exhibits a maximum intensity along the direction perpendicular to the field, which is due to the term $|\widetilde{M}_s|^2 \sin^2\theta$ in $d \Sigma_{\mathrm{res}} / d \Omega$. Reducing the field to $196 \, \mathrm{mT}$ results in the emergence of transversal spin-misalignment fluctuations (in addition to the $|\widetilde{M}_z|^2 \sin^2\theta$ contribution), which give rise to angular anisotropies with maxima along the horizontal direction and (roughly) along the detector diagonals [compare the expressions for both response functions in Eqs.~(\ref{rhdefperp}) and (\ref{rmdefperp})]. This is clearly revealed by inspection of the spin-misalignment SANS at $196 \, \mathrm{mT}$ [Fig.~\ref{fig1}(c)], which shows (i)~a weak clover-leaf-type anisotropy and (ii)~an elliptical elongation along the field direction. The scattered dots (speckles) at the outskirts of the cross section in Fig.~\ref{fig1}(c) indicate the presence of a (nearly) $q$-vector independent small random error in the data, which we estimate to be around $\pm 30 \, \mathrm{cm}^{-1} \mathrm{sr}^{-1}$. Note that azimuthal averaging, performed in Ref.~\onlinecite{honecker2013}, smoothes this error out. It has bigger impact on the present third-order effect analysis, which is based on the subtraction of the cross-section values along only two (vertical and horizontal) directions on the detector [Eq.~(\ref{eq:deltasigma})]. The origin of the clover-leaf anisotropy is related to the dipolar stray fields that are due to the jumps in $M_s$ at particle-matrix interfaces (as in the case of nanoporous Fe, see Fig.~\ref{fig3}). These stray fields decorate the Fe nanoparticles via the exertion of a magnetic torque on the magnetic moments of the matrix~\cite{honecker2013,bischof07,michels06prb}. 

The analysis in Ref.~\onlinecite{honecker2013} yields a value for the exchange stiffness of $A = 4.7 \pm 0.9 \, \mathrm{pJ/m}$. Also, the fitted values of $S_H(q)$ are many orders of magnitude smaller than $S_M(q)$ (see Fig.~5 in Ref.~\onlinecite{honecker2013}). This means that magnetostatic effects (due to the small spatial variation of the saturation magnetization) are dominating over the anisotropy ones (due to the small spatial variation of the anisotropy constant). Thus, we can ignore the latter and assume $\kappa = 0$ in Eq.~(\ref{eq:sigmaperpthree}), which makes it independent of the function $g_{\mathrm{A}}$. Finally, the $q$-dependence of $S_M$ seems to be better described by an exponential function rather than a Gaussian (Gaussian spatial profile of the defects remains Gaussian in $q$-space). This was verified by us using the numerical data of Ref.~\onlinecite{honecker2013}. The exponential $S_M(q)$ corresponds to a Lorentzian-squared defect profile in real space, such as the following model for $I(\mathbf{r})$:
\begin{equation}
\label{defectsModel}
I(\mathbf{r}) = \sum_i \frac{a_i}{(1 + |\mathbf{r} - \mathbf{r}_i|^2/s^2)^2} - const ,
\end{equation}
where $a_i \ll 1$ and $\mathbf{r}_i$ are, respectively, the random amplitudes and positions of the defects, and the summation runs over the sample volume. The value of the additive constant is chosen to ensure that $\langle I(\mathbf{r}) \rangle = 0$, which is always possible for the considered defect profile.

Substituting the first-order micromagnetic solutions for $\widetilde{M}_x^{(1)}$ and $\widetilde{M}_y^{(1)}$ from Ref.~\onlinecite{metmi2015} into Eq.~(\ref{thirdorderintermediate}), passing from a summation to an integration, and assuming $\kappa = 0$ yields the following expression:
\begin{widetext}
\begin{eqnarray}
\label{eq:gms}
 g_{\mathrm{MS}}=\frac{v}{(2\pi)^3}\left\langle\widetilde{I}(\mathbf{q})\iiint \frac{(x_{\mathbf{q}-\mathbf{q}^\prime} x_{\mathbf{q}^\prime}+y_{\mathbf{q}-\mathbf{q}^\prime} y_{\mathbf{q}^\prime})z_{\mathbf{q}-\mathbf{q}^\prime} z_{\mathbf{q}^\prime}\widetilde{I}(\mathbf{q}-\mathbf{q}^\prime)\widetilde{I}(\mathbf{q}^\prime)}{2(h_{\mathbf{q}-\mathbf{q}^\prime}+x^2_{\mathbf{q}-\mathbf{q}^\prime}+y^2_{\mathbf{q}-\mathbf{q}^\prime})(h_{\mathbf{q}^\prime}+x^2_{\mathbf{q}^\prime}+y^2_{\mathbf{q}^\prime})} d^3\mathbf{q}^\prime\right\rangle ,
\end{eqnarray} 
\end{widetext}
where $\{x_\mathbf{q},y_\mathbf{q},z_\mathbf{q}\} = \mathbf{q}/q$, $v=\iiint 1/(1+r^2/s^2)^2 d^3\mathbf{r}=\pi^2 s^3$ is the volume of the single defect to make the result dimensionless, and $h_\mathbf{q} = h + l_M^2 |\mathbf{q}|^2$. The integral results from the convolution and the angular brackets correspond to the directional (over different representative volume orientations) and the ensemble [over different realizations of the random process for $I(\mathbf{r})$] averaging. Unlike the consideration of $g_{MS}$ in Ref.~\onlinecite{metmi2015}, the Eq.~(\ref{eq:gms}) does not assume a specific defect model $I(\mathbf{r})$. Inserting the Fourier transform $\widetilde{I}(\mathbf{q})$ of Eq.~(\ref{defectsModel}) and performing the averaging results in a slightly more complicated expression, which was used in the actual computations (see the Supplemental Mathematica file~\cite{suppmathfile}).
\begin{figure}[tb!]
\centering{\includegraphics[width=1.0\columnwidth]{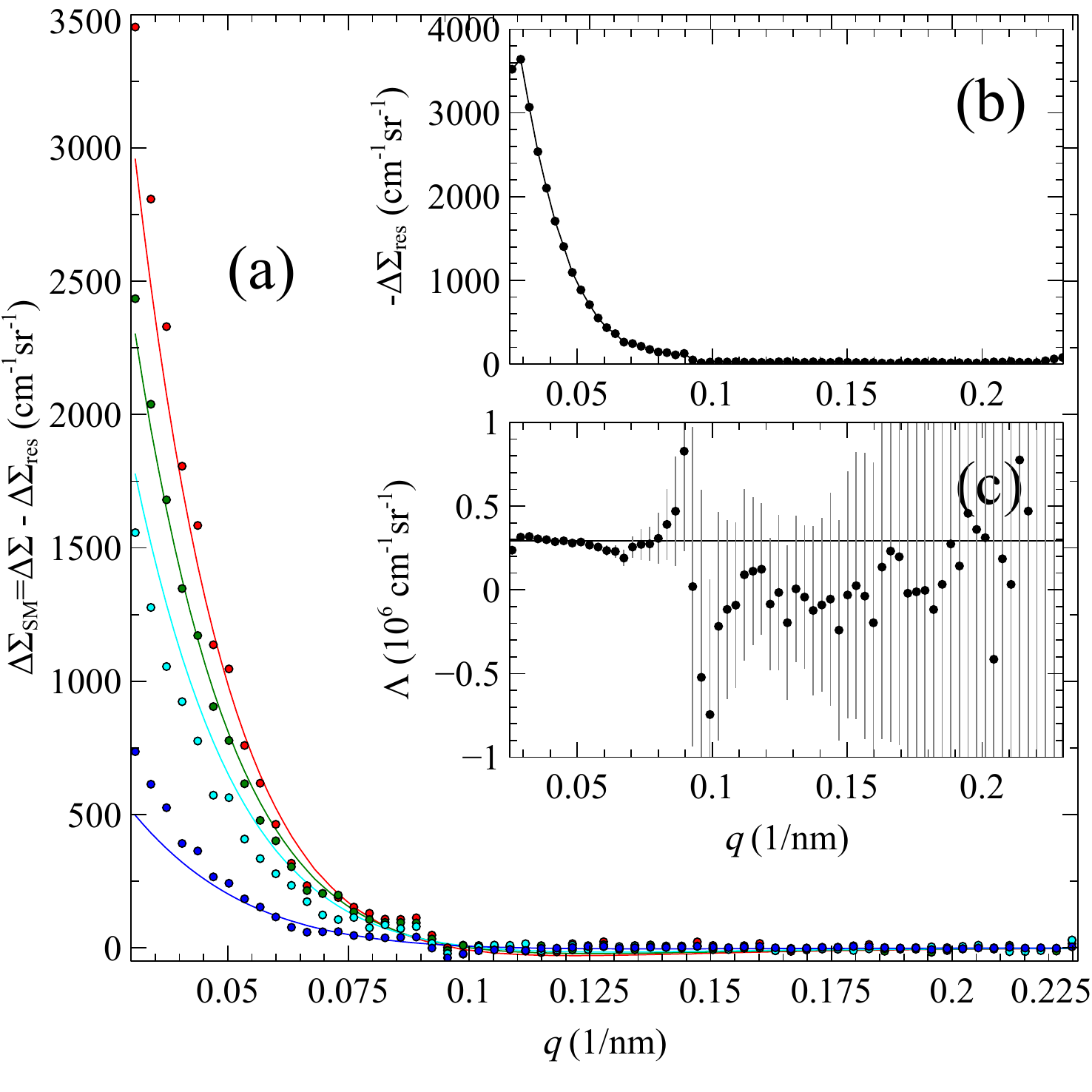}}
\vspace{-0.5cm}
\caption{Spin-misalignment SANS cross-section differences of NANOPERM [$(\mathrm{Fe}_{0.985}\mathrm{Co}_{0.015})_{89}\mathrm{Zr}_7\mathrm{B}_3$] at selected applied magnetic fields. Field values from bottom to top---$1270 \, \mathrm{mT}$, $312 \, \mathrm{mT}$, $196 \, \mathrm{mT}$, $103 \, \mathrm{mT}$. The points are the experimental data, the solid lines in (a) are computed from Eq.~(\ref{eq:ordIII}) with $\Lambda = \Lambda_0 = 0.29 \times 10^{6} \, \mathrm{cm}^{-1}\mathrm{sr}^{-1}$ and $g_{\mathrm{MS}}$, corresponding to the Lorentzian-squared defects (Eq.~(\ref{defectsModel}) with $s = 19.5 \, \mathrm{nm}$). Insets show the fitted dependencies of $\Delta\Sigma_\mathrm{res}(q)$ and $\Lambda(q)$. The horizontal line in (c) is $\Lambda = \Lambda_0$.}
\label{fig2}
\end{figure}

The main problem in analyzing experimental data, measured at finite fields, is to find and subtract the field-independent residual SANS cross section. Even though at a large magnetic field the average magnetization is close to saturation, there are still many local fluctuations, which can be detected by the extremely sensitive SANS technique. The assumption of magnetic saturation at the largest of the experimentally achievable magnetic field can be avoided by analyzing a combination of the cross-section values similar to Eq.~(\ref{eq:deltasigma}), but constructed from the total (nuclear and magnetic) cross-section data as opposed to the spin-misalignment part only:
\begin{eqnarray}
\label{eq:ordIII}
\Delta\Sigma & = &\left. \frac{d \Sigma}{d \Omega} \right|_{\theta = 0} - 2 \left. \frac{d \Sigma}{d \Omega}\right|_{\theta = \pi/2} \nonumber \\
  & = & \Delta\Sigma_\mathrm{res}(q) + \Lambda \, g_{\mathrm{MS}}(q s, H_i/M_s, l_M/s).  
 \end{eqnarray}
For the same reasons as before, $\Delta\Sigma$ contains no second-order term. It consists only of the magnetic-field-independent residual contribution $\Delta\Sigma_\mathrm{res} = \frac{8 \pi^3}{V} [|\widetilde{N}|^2(\{0,0,q\}) - 2 |\widetilde{N}|^2(\{0,q,0\}) - 2 b_H^2 |\widetilde{M}_s|^2(\{0,q,0\})]$ and (if fourth and higher-order terms are neglected) of the third-order term. In the case of our material with $S_H \ll S_M$ the latter is proportional to $g_{\mathrm{MS}}$ with a field and $q$-independent scaling parameter $\Lambda$. Because the large nuclear and residual cross sections are subtracted twice in Eq.~(\ref{eq:ordIII}), both $\Delta\Sigma$ and $\Delta\Sigma_\mathrm{res}$ are negative, while $\Lambda$ is positive.

Linearity of Eq.~(\ref{eq:ordIII}) (in $g_{\mathrm{MS}}$) suggests the possibility of fitting the field dependence of the experimental $\Delta\Sigma$ at each $q$ as a function of the computed value of $g_{\mathrm{MS}}$ using linear regression. The only remaining parameter is the size of the defects $s$, which can be adjusted iteratively to minimize the total error of the fit. This procedure results in the best-fit value of $s = 19.5 \, \mathrm{nm}$ and in the corresponding $\Delta\Sigma_\mathrm{res}(q)$ and $\Lambda(q)$ dependencies as shown in Fig.~\ref{fig2}(b) and Fig.~\ref{fig2}(c). The $s$-value agrees very well with the nominal particle size of $15 \pm 2 \, \mathrm{nm}$ of the alloy.

It is important to note that, while theoretically the value of $\Lambda$ should be independent of $q$, this dependence was allowed during the fitting procedure. If the specific choice of the defect profile and the resulting $g_{\mathrm{MS}}$ is viable, a $q$-independent $\Lambda$-value should come as the result of the fit. As one can see from Fig.~\ref{fig2}(c), this is indeed the case. The error bars were computed using a Monte-Carlo procedure by adding a random $\pm 30 \, \mathrm{cm}^{-1}\mathrm{sr}^{-1}$ contribution to the measured $d \Sigma / d \Omega$-values and computing the standard deviation of the resulting $\Lambda$ at each $q$ across many realizations of this random process. The above value of the assumed absolute error has been estimated from the scatter of the data at the outskirts of the cross sections (at $q > 0.1 \, \mathrm{nm}^{-1}$) shown in Fig.~\ref{fig1}(c). For $q \lesssim 0.1 \, \mathrm{nm}^{-1}$, the $\Lambda(q)$ assume a nearly constant value of $\Lambda = \Lambda_0 = 0.29 \times 10^{6}\, \mathrm{cm}^{-1}\mathrm{sr}^{-1}$, shown by the horizontal line. The amplification of the error at larger $q$ is due to the small value of the cross section at these scattering vectors and, precisely for this reason, is, probably, of no relevance. This is corroborated by the theoretical lines in Fig.~\ref{fig2}(a), which are plotted according to Eq.~(\ref{eq:ordIII}) using the fixed $q$-independent value of $\Lambda = \Lambda_0$. The solid lines fit the experimental data reasonably well in the approach-to-saturation regime.

The fit in Fig.~\ref{fig2}(a) is good, but not perfect. The probable reason is the sensitivity of the cross-section difference to the details of the shape of the inhomogeneities. That is why we have carefully evaluated the second-order SANS cross section of this sample from Ref.~\onlinecite{honecker2013} and fitted $S_M(q)$ by the Fourier image of a Lorentzian-squared defect profile. While such a fit describes the $q < 0.1 \, \mathrm {nm}^{-1}$ region of $S_M(q)$ well, it exhibits discrepancies at larger $q$. These discrepancies are irrelevant to the second-order SANS theory and impact the cross section at large $q$-values only, where it is very small. However, because of the convolution in the third-order difference function Eq.~(\ref{eq:gms}), they influence the $\Delta\Sigma$ at smaller $q$ as well. It means that the interpretation of the third-order SANS cross-section differences is much more demanding on the precision of the defect model and can be a valuable tool for gaining additional insights about the shape of the inhomogeneities in the material.

While the cross-section values themselves are strictly positive, the $\Delta\Sigma$ may assume negative values for some $q$ (see Fig.~5 in Ref.~\onlinecite{metmi2015}). The $\Delta\Sigma_{\mathrm{SM}}$ in Fig.~\ref{fig2}(a) also have some visibly negative points on the lower curves.

We would like to remind that the microstructure of NANOPERM is very different from the one used in the simulations. The simulated system has nanopores instead of nanocrystallites as in the NANOPERM sample. That is why a direct comparison between Fig.~\ref{fig3} and Fig.~\ref{fig1} is not very useful. It is generally a very difficult problem to simulate a realistic random nanostructured system on a scale necessary for computing the SANS cross section. While both the numerical simulations and the analytical theory we use are approximate, their approximations are different. The simulation is not limited by the series expansion, but it is limited by the relatively small statistics of the material nanostructure, represented in the simulation volume. On the other hand, the analytical theory includes the full statistical averaging over an infinite volume, but is limited by the third-order terms in the material inhomogeneities amplitude. Yet, despite these shortcomings and differences between the simulation and theory both reveal the presence of the third-order effect.

An applied field of $\mu_0 H_0 = 1270 \, \mathrm{mT}$ seems to be rather large, and polarizes the material close to the saturation, as can be seen from the hysteresis loop in Fig.~\ref{fig1}(d). Yet, the corresponding third-order spin-misalignment SANS cross-section difference, shown as the bottom line in Fig.~\ref{fig2}(a), is far from zero (as it should be in the case of infinite external field). This is another illustration of the extreme sensitivity of SANS to the inhomogeneities of the sample's magnetization, by far exceeding that of traditional magnetometry.

One of the completely new possibilities which are opened-up by the observation of the third-order effect is the ability to extract the third statistical moment of the defects-magnitude distribution. This follows since $g_{\mathrm{MS}} \sim <a_i^3>$ [compare Eqs.~(\ref{defectsModel}) and (\ref{eq:gms})]. The third moment measures the skewness in the statistical distribution of the nanocrystallite sizes (if they are made of the same material). The skewness is zero if the distribution is symmetric around its center (like e.g.\ for a Gaussian). It will be interesting in future studies to explore the evolution of the skewness, which is expected to develop during the various stages of nanocrystallization.

\section{Conclusions}

We have demonstrated both numerically and experimentally the existence of the theoretically predicted third-order effect in the magnetic SANS cross section, which cannot in principle be accounted for by the second-order SANS theory. The model of Ref.~\onlinecite{metmi2015} is extended and the resulting expressions describe both the third-order effect and its field dependence well. Because of the inherent convolution in $\mathbf{q}$-space, the third-order effect is much more sensitive to the details of the defect profile as compared to the second-order SANS theory. We have provided here the general expression for its field and scattering-vector dependence, suitable for an arbitrary spatial-inhomogeneity profile, and used a Lorentzian-squared profile in our analysis. Analyzing the data with the help of the third-order SANS theory does not require new SANS measurements. It can make SANS an even more valuable and powerful tool for the microstructure analysis of magnetic materials.

\begin{acknowledgments}
We would like to thank Sergey Erokhin and Dmitry Berkov (General Numerics Research Lab, Jena, Germany) for providing the micromagnetic simulation results. KLM acknowledges the support of the Russian Science Foundation under the project RSF~16-11-10349.
\end{acknowledgments}

\bibliography{BIB,BIB1}

\end{document}